\documentclass[preprint,amsmath,amssymb]{revtex4}
\usepackage{amssymb}
\usepackage{graphicx}
\usepackage{dcolumn}
\usepackage{bm}

\makeatletter

\begin{document}

\title{Structures, Electronic Properties, Spectroscopies and Hexagonal Monolayer Phase of a Family of Unconventional Fullerenes C$_{64}$X$_{4}$ ($X=H,F,Cl,Br$)}
\author{Qing-Bo Yan}
\author{Qing-Rong Zheng}
\author{Gang Su}
\email[Author to whom correspondence should be addressed. ]{Email:
gsu@gucas.ac.cn} \affiliation{{College of Physical Sciences,
Graduate University of Chinese Academy of Sciences, P.O. Box 4588,
Beijing 100049, China}}

\begin{abstract}
A systematic first-principles study within density functional theory
on the geometrical structures and electronic properties of
unconventional fullerene C$_{64}$ and its derivatives
C$_{64}$X$_{4}$ ($X=H,F,Cl,Br$) has been performed. By searching
through all 3465 isomers of C$_{64}$, the ground state of C$_{64}$
is found to be spherical shape with D$_{2}$ symmetry, which differs
from the parent cage of the recently synthesized C$_{64}$H$_{4}$
that is pear-shaped with C$_{3v}$ symmetry. We found that the
addition of the halogen atoms like $F,Cl,Br$ to the
pentagon-pentagon fusion vertex of C$_{64}$ cage could enhance the
stability, forming the unconventional fullerenes C$_{64}$X$_{4}$.
The Mulliken charge populations, LUMO-HOMO gap energies and density
of states are calculated, showing that different halogen atoms added
to C$_{64}$ will cause remarkably different charge populations of
the C$_{64}$X$_{4}$ molecule; the chemical deriving could enlarge
the energy gaps and affect the electronic structures distinctly. It
is unveiled that C$_{64}$F$_{4}$ is even more stable than
C$_{64}$H$_{4}$, as the C-X bond energy of the former is higher than
that of the latter. The computed spectra of C$_{64}$H$_{4}$
molecules agree well with the experimental data; the IR, Raman, NMR
spectra of C$_{64}$X$_{4}$ ($X=F,Cl,Br$) are also calculated to
stimulate further experimental investigations. Finally, it is
uncovered by total energy calculations that C$_{64}$X$_{4}$ could
form a stable hexagonal monolayer.
\end{abstract}

\maketitle

Fullerenes are closed-cage carbon molecules of 12 pentagons and
several hexagons. Since discovery of Buckminster fullerene
C$_{60}$\cite{c60discover} in 1985, intensive theoretical and
experimental works have been done to study the structures and
properties of fullerenes. The conventional fullerenes such as C$_{60}$, C$_{70}$,
C$_{70+2n},n=1,2,\ldots$ are satisfying the empirical isolated
pentagon rule (IPR)\cite{IPR1,IPR2}, which states that the most
stable fullerenes are those in which every pentagon is surrounded by
five hexagons, ensuring the minimum curvature of the fullerene cage
(i.e., the minimum strain)\cite{minimum_curvaure}. Fullerenes with
other number of carbons are impossible to have all the pentagons
isolated in the carbon cage, thus they are called unconventional fullerenes or
non-IPR fullerenes, which
are believed to be labile and difficult to synthesize. In fact, most
of the several stable fullerene isomers\cite{stablefuller} which had
been isolated are satisfying IPR. For applications, fullerenes and
their derivatives are potential new agents and materials for
molecular electronics, nanoprobes, superconductors, and nonlinear
optics. Thus if IPR could be efficiently violated in some ways, the
bank of stable fullerenes for practical applications will be much
enriched.

In the recent years, it has been uncovered that some non-IPR
fullerenes may be stabilized through metallic endohedral
(Sc$_{2}$@C$_{66}$\cite{endohedral_CRWang},
A$_{x}$Sc$_{3-x}$N@C$_{68}$\cite{endohedral_Stevenson},
La$_{2}$@C$_{72}$\cite{endohedral_Kato}, etc.). The other effort and
progress have also been made in chemical deriving. In 2004, Xie
\textit{et al.}\cite{Xie} had successfully synthesized a new
exohedral chemical deriving non-IPR D$_{5h}$ fullerene[50] (say,
C$_{50}$Cl$_{10}$, with 10 chlorine atoms added to the
pentagon-pentagon vertex fusions) in milligram by a graphite
arc-discharge process modified by introducing a small amount of
carbon tetrachloride (CCl$_{4}$) in the helium atmosphere. Later,
another new chemical deriving non-IPR fullerene[64] (namely,
C$_{64}$H$_{4}$, with 4 hydrogen atoms added to the vertex of a
triplet directly fused pentagons) has been successfully prepared by
Wang \textit{et al.}\cite{Wang} in milligram with similar method. As
CCl$_{4}$ was added to provide chlorine source in the case of
C$_{50}$Cl$_{10}$, methane (CH$_{4}$) was introduced to provide
hydrogen for C$_{64}$H$_{4}$. Both C$_{50}$Cl$_{10}$ and
C$_{64}$H$_{4}$ are stabilized by the chemical modification on the
pentagon-pentagon fusion vertex. This clue hints a promising general
method to stabilize non-IPR fullerenes. While an efficient and
controllable method to prepare the metallic endohedral fullerenes
with macroscopic quantity is not yet matured, the chemical deriving
might open a useful way to obtain stable non-IPR fullerenes.

Moreover, many questions are still unclear: What is the ground state
of C$_{64}$? How about the electronic properties? Since the adding
atoms in C$_{50}$Cl$_{10}$ are Cl, what will happen if the H atoms
in C$_{64}$H$_{4}$ are replaced by other atoms, for instance, F, Cl,
Br? Here, by first-principles density functional calculations, we
shall address these issues, and report intensive studies on the
geometrical structures, electronic properties and spectroscopies of
C$_{64}$ and its derivatives C$_{64}$X$_{4}$ ($X=H,F,Cl,Br$). As it
is uncovered that C$_{50}$Cl$_{10}$ molecules could form a hexagonal
monolayer \cite{C50Cl10monolayer}, we also investigate the
possibility that C$_{64}$X$_{4}$ could form a hexagonal monolayer.
Our present study would deepen understanding of the properties of the recently isolated fulleride
C$_{64}$H$_{4}$, and could also stimulate further experimental efforts
on C$_{64}$X$_{4}$ ($X=F,Cl,Br$).


We have generated all of the 3465 topologically distinct isomers of
C$_{64}$ cage by spiral algorithm\cite{spiral}, and sorted them by
the corresponding pentagon-pentagon adjacency (PPA) numbers.
(Definition: one distinct C-C bond shared by two adjacent pentagons
in fullerenes is counted as one PPA.) Then, we have
calculated the $C_{64}$ isomers with small PPA numbers ($N_{\mathtt{PPA}}%
\mathtt{=2,3,4}$) by the semiempirical MNDO method\cite{MNDO}. (see
supporting materials.) In Gaussian-like terminology, that would
be called MNDO//MNDO calculations, in which \emph{B}//\emph{A}
represents \emph{B} level energy calculation based on the geometry
optimized on \emph{A} level. It had been proved that the
semiempirical MNDO method could reproduce the ab initio or density
functional relative energies of fullerene isomers with useful high
accuracy\cite{semiempirical, C40PAPR}.

Among them, nine isomers of C$_{64}$ with the lowest energies (Fig.
1) were chosen, and their derivatives were generated. Using the
density functional theory (DFT)\cite{Hohenberg-Kohn}, we have
calculated geometries and relative energies of above nine isomers
and their derivatives. In particular, we have invoked the B3LYP
functional, which combines the Becke three parameter non-local
hybrid exchange potential\cite{B3} and the nonlocal correlation
functional of Lee-Yang-Parr\cite{LYP}. The basis set 6-31G$^{*}$ was
used in the B3LYP calculations. The B3LYP/6-31G$^{*}$ calculations
also give the LUMO-HOMO gaps. The harmonic vibrational frequencies
and nuclear magnetic shielding tensors were calculated at the
B3LYP/6-31G$^{*}$ and GIAO-B3LYP/6-31G$^{*}$ level, respectively.

\begin{figure}[tb]
\includegraphics[width=3.5in, keepaspectratio]{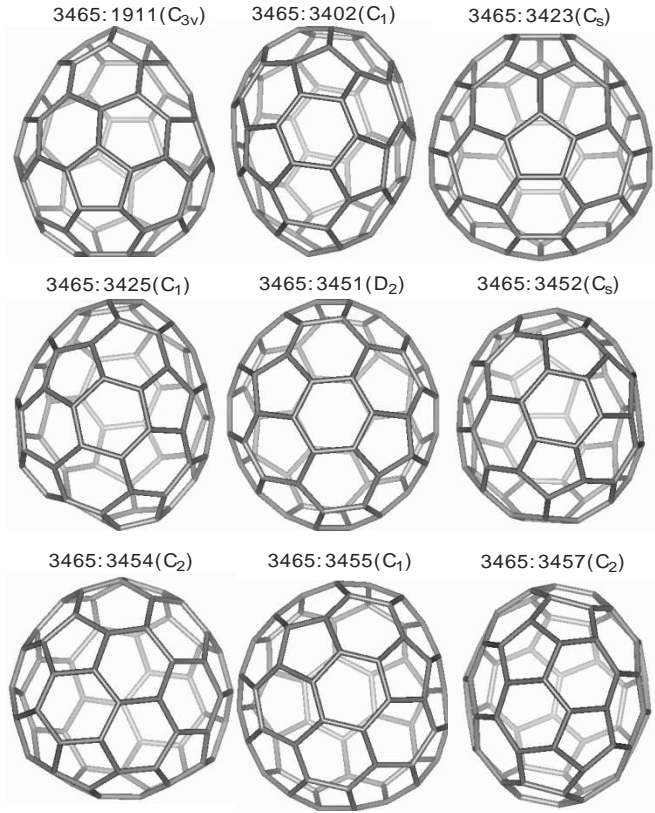}
\caption{The 9 lowest-energy isomers of C$_{64}$ obtained at MNDO. }
\label{c64lowest9}
\end{figure}

All of the calculations are completed by Gaussian 03\cite%
{Gaussian03} except that the density of states (DOS) and monolayer
total energies of C$_{64}$X$_{4}$ are calculated by means of the
ABINIT package\cite{ABINIT}. This package is coded within the DFT
framework based on pseudopotentials and plane waves. Particularly,
in our calculations, Troullier-Martins norm conserving
pseudopotentials\cite{NCPP} and the Teter
parametrization\cite{Teter} of the Ceperley-Alder
exchange-correlation potential were used.



Fig. 1 shows shapes and symmetries of the nine lowest-energy isomers
of C$_{64}$ obtained at MNDO level. They have been labeled as
$3465:\emph{n}$ , where \emph{n} denotes the isomer number in spiral
nomenclature\cite{spiral}. We reoptimized the geometries and
recalculated their relative energies by B3LYP/6-31G$^{*}$ method,
where the LUMO-HOMO gap energies are obtained. The details are
listed in Table \ref{Table1}.

\begin{table}[tb]
\caption{MNDO//MNDO, B3LYP/6-31G$^{*}$//B3LYP/6-31G$^{*}$ relative
energies (in eV), LUMO-HOMO gap energies (E$_{\texttt{gap}}$, in eV)
of the nine lowest-energy C$_{64}$ isomers}
\label{Table1}\setlength{\textfloatsep}{0.5cm}
\begin{tabular}{cccccccccccccc}
 \hline\hline
  &   & & & \multicolumn{2}{c}{Relative Energy} & & &\\
\cline{5-6}\cline{9-11} Isomer\footnotemark[1] &  Symm & N$_{\texttt{PPA}}\footnotemark[2]$ & \hspace{1mm}N$_{\texttt{PPFV}}$\footnotemark[3] & \hspace{1mm}MNDO\footnotemark[4]\hspace{2mm} & \hspace{1mm}B3LYP\footnotemark[5]\hspace{1mm} & \hspace{2mm}E$_{\texttt{gap}}$ \hspace{3mm}&&\\
\hline
1 (1911) & C$_{\texttt{3v}}$ & 3 & 4 & 1.0979 [4] & 1.2403 [6] & 1.0879 & & \\
2 (3402) & C$_{\texttt{1}}$ & 3 & 6 & 1.4144 [8] & 1.4706 [9] & 1.4245 & & \\
3 (3423) & C$_{\texttt{s}}$ & 3 & 6 & 1.1597 [6] & 1.2156 [5] & 1.3845 & & \\
4 (3425) & C$_{\texttt{1}}$ & 3 & 6 & 1.4319 [9] & 1.2995 [7] & 1.2408 & & \\
5 (3451) & D$_{\texttt{2}}$ & 2 & 4 & 0.0000 [1] & 0.0000 [1] & 2.2327 & & \\
6 (3452) & C$_{\texttt{s}}$ & 2 & 4 & 0.2600 [2] & 0.2987 [2] & 2.0482 & & \\
7 (3454) & C$_{\texttt{2}}$ & 3 & 6 & 1.3619 [7] & 1.4330 [8] & 1.6311 & & \\
8 (3455) & C$_{\texttt{1}}$ & 3 & 6 & 1.1178 [5] & 1.1116 [4] & 1.9369 & & \\
9 (3457) & C$_{\texttt{2}}$ & 2 & 4 & 0.4807 [3] & 0.5563 [3] & 1.8547 & & \\
\hline\hline
\end{tabular}
\footnotetext[1]{Labeled as m(n) form, where (n) was the numbering
in spiral nomenclature\cite{spiral}.} \footnotetext[2]{Number of
pentagon-pentagon adjacency (PPA).} \footnotetext[3]{Number of
pentagon-pentagon fusion vertex (PPFV).} \footnotetext[4]{[n]
indicates the stability order on MNDO level.} \footnotetext[5]{[n]
indicates the stability order on B3LYP/6-31G$^{*}$ level.}
\end{table}

As seen in Table \ref{Table1}, the isomer 3451 (numbering in spiral
nomenclature\cite{spiral}) is the ground state of C$_{64}$, followed
by the isomer 3452 and 3457. The isomer 3451 is nearly spherical
shaped, with D$_{\texttt{2}}$ symmetry, while the isomer 3452 and
3457 bear C$_{\texttt{s}}$ and C$_{\texttt{2}}$ symmetry,
respectively. These three lowest-energy isomers have two pairs of
adjacent pentagons (counted as 2-PPA), and the rest isomers with
higher energies all have three couples of adjacent pentagons
(counted as 3-PPA) except that the isomer 1911, has a triplet of
fused pentagons (also counted as 3-PPA). This pear-shaped C$_{3v}$
symmetry isomer 1911, which is just the carbon cage parent of the
synthesized C$_{64}$H$_{4}$ by Wang \emph{et al}.\cite{Wang},
however, is not apparently favored in energy when it is compared
with other eight isomers. The ground state isomer 3451 also has the
largest LUMO-HOMO gap energy, followed by the isomer 3457, whereas
that of the isomer 1911 is the smallest in the nine.

It may be worth to note that the stability order of the nine isomers
obtained by B3LYP/6-31G$^{*}$ calculations is similar, but not
exactly the same to that by MNDO. Both methods give the same order
for the most stable structures (the isomer 3451, 3452, 3457).


The pentagon-pentagon fusion vertex (PPFV) are usually considered to
be the most active sites of the fullerene\cite{C50_XinLu}, and the
syntheses of stable C$_{50}$Cl$_{10}$ and C$_{64}$H$_{4}$ also
clearly evidence that the chemical modification on the PPFV could
effectively release the large steric strain of non-IPR fullerenes.
Here we attach H and Cl atoms to the PPFV of above nine C$_{64}$
isomers to construct C$_{64}$ fullerene derivatives. As listed in
Table \ref{Table1}, isomers 1911, 3451, 3452, 3457 have 4 PPFV,
while others have 6 PPFV. Thus, the corresponding fullerene
derivatives are C$_{64}$X$_{4}$ and C$_{64}$X$_{6}$ ($X=H$ or $Cl$),
respectively. Since the number of the added X atom is not fixed, the
relative energy is not suitable for evaluating the relative
stability. In order to discuss the relative stability of
C$_{64}$X$_{n}$, we define C-X bond energy $\Delta{\texttt{E}}$ as
follows \cite{Delta_E}:
$$\Delta{E}=\frac{1}{n}(E_{\texttt{C}_{64}}+nE_{\texttt{X}}-E_{\texttt{C}_{64}\texttt{X}_{n}})$$
The higher the C-X bond energy $\Delta{\texttt{E}}$, the greater the
stability of C$_{64}$X$_{n}$.

\begin{table}[tb]
\caption{B3LYP/6-31G$^{*}$//B3LYP/6-31G$^{*}$ bond energies
$\Delta{\texttt{E}}$ (in eV), LUMO-HOMO gap energies
(E$_{\texttt{gap}}$, in eV) of the derivatives of nine lowest-energy
C$_{64}$ isomers} \label{Table2}\setlength{\textfloatsep}{0.5cm}
\begin{tabular}{ccccccccccc}
\hline\hline
  & & \multicolumn{2}{c}{C$_{64}$H$_{n}$} & &\multicolumn{2}{c}{C$_{64}$Cl$_{n}$} \\
\cline{3-4}\cline{6-7}
  Isomer & $n$\footnotemark[1] \hspace{1mm}&  $\Delta$E(B3LYP)\hspace{1mm} & \hspace{4mm}E$_{\texttt{gap}}$\hspace{2mm} &  & \hspace{1mm}$\Delta$E(B3LYP)\hspace{1mm} & \hspace{4mm}E$_{\texttt{gap}}$\hspace{2mm} &\\
\hline
1(1911) &  4 & 3.9019[1] & 2.6697 &&  2.6261[1] & 2.6923 & \\
2(3402) &  6 & 3.1113[4] & 0.9611 &&  1.8532[4] & 0.8963 & \\
3(3423) &  6 & 3.0093[7] & 0.9793 &&  1.7513[7] & 1.0068 & \\
4(3425) &  6 & 2.9186[8] & 0.6833 &&  1.7008[8] & 0.7959 & \\
5(3451) &  4 & 2.8985[9] & 1.3557 &&  1.6392[9] & 1.2398 & \\
6(3452) &  4 & 3.0153[6] & 1.3968 &&  1.7628[6] & 1.3685 & \\
7(3454) &  6 & 3.1407[3] & 1.7949 &&  1.8827[3] & 1.7374 & \\
8(3455) &  6 & 3.0754[5] & 1.2651 &&  1.8214[5] & 1.2441 & \\
9(3457) &  4 & 3.1920[2] & 1.9108 &&  1.9412[2] & 1.8623 & \\
\hline\hline
\end{tabular}
\footnotetext[1]{the number of added H or Cl atoms, i. e., the
number of pentagon-pentagon fusion vertex (PPFV) for each isomer
according to the method we generated the derivatives.}
\end{table}

Table \ref{Table2} lists the B3LYP/6-31G$^{*}$//B3LYP/6-31G$^{*}$
C-X bond energy $\Delta{\texttt{E}}$ and relative stability,
LUMO-HOMO gap energies of the hydro- and chloro- derivatives of nine
lowest-energy C$_{64}$ isomers such as C$_{64}$X$_{n}$ ($X=H$ or
$Cl$, $n=4$ or $6$). We still use the spiral nomenclature to label
the generated fullerene derivatives.

The stability order is changed after H atoms are attached to the
C$_{64}$ isomers. As expected, the C$_{64}$H$_{4}$ 1911, with a
remarkably largest C-H bond energy, turns out to be the most stable
in all C$_{64}$H$_{n}$ derivatives. The following ones are
C$_{64}$H$_{4}$ 3457 and C$_{64}$H$_{6}$ 3454, both having C$_{2}$
symmetry. Derivative 3451, which has been evolved from the ground
state of C$_{64}$ isomers, becomes the most instable among the nine
derivatives.

Corresponding to its remarkable larger C-X bond energy
$\Delta{\texttt{E}}$, the C$_{64}$H$_{4}$ 1911 has the largest
LUMO-HOMO gap energy, 2.6697 eV, while that of its parent C$_{64}$
1911 is only 1.0879 eV. The LUMO-HOMO gap energies of others are
also shifted, implying that the chemical deriving on PPFV could
affect the electronic structures of non-IPR fullerenes.

Obviously, the chemical deriving on PPFV could indeed enhance the
stability of the non-IPR fullerenes in the case from C$_{64}$ 1911
to C$_{64}$H$_{4}$ 1911. However, the effects are different for
different isomers. It seems that the vertex of the triplet fused
pentagons are rather more favorable for hydrogen addition than other
types of vertex, which is consistent with the suggestion by Liu
\emph{et al}. \cite{exhydrogenate_order}.

The effects of Cl addition are similar to that of the H addition.
After Cl atoms attached to the C$_{64}$ isomers, the stability order
is changed, and the new order is entirely the same as that of the H
addition. C$_{64}$Cl$_{4}$ 1911 also becomes the most stable with a
remarkable larger C-Cl bond energy. The LUMO-HOMO gap energies are
changed with the same manner too. C$_{64}$Cl$_{4}$ 1911 also has the
highest LUMO-HOMO gap energy, 2.6923 eV, in all C$_{64}$Cl$_{n}$
derivatives. These give us a clue that haloid atoms may be
appropriate for the chemical deriving on the PPFV of non-IPR
fullerenes to stabilize the structure.


Now we attach H, F, Cl, Br atoms to the PPFV of the isomer C$_{64}$
1911 to construct C$_{64}$X$_{4}$ ($X=H, F, Cl, Br$) fullerene
derivatives and, to study their structural and electronic
properties. Fig. 2 displays the geometrical structures of
C$_{64}$X$_{4}$ ($X=H, F, Cl, Br$). These molecules like pears in
shape with four short hairs on its tapering top or odd pears with
four stalks. The tapering top of C$_{64}$ cage is composed of a
triplet of directly fused pentagons. Its strain is obviously high
due to its high surface curvature, which is a key factor for its
instability. The length of C-X bond along the $C_{3}$ axis in
C$_{64}$X$_{4}$ ($X=H, F, Cl, Br$) is about $1.09$, $1.36$, $1.77$,
$1.93$ {\AA}, respectively, while the length of other C-X ($X=H, F,
Cl, Br$) bonds around the $C_{3}$ axis is about $1.10$, $1.38$,
$1.82$, $1.99$ {\AA}, respectively. The bond lengths of C-X bonds
along the $C_{3}$ axis are nearly the same as the experimental value
of C-X ($X=H, Cl, Br$) bond length in methane (CH$_{4}$), CF$_{4}$,
chloroform (CCl$_{4}$) or CBr$_{4}$, while the other bonds around
the $C_{3}$ axis are somewhat longer. For a comparison, we have
tried to put $H, F, Cl$ or $Br$ atom along the $C_{3}$ axis into the
inner of C$_{64}$ cage, and found that the calculated energies are
higher than those when these atoms are added to the outer of the
cage, suggesting that the exohedral fullerenes C$_{64}$X$_{4}$
($X=H, F, Cl, Br$) are more stable.

\begin{figure}[tb]
\includegraphics[width=3.4in, keepaspectratio]{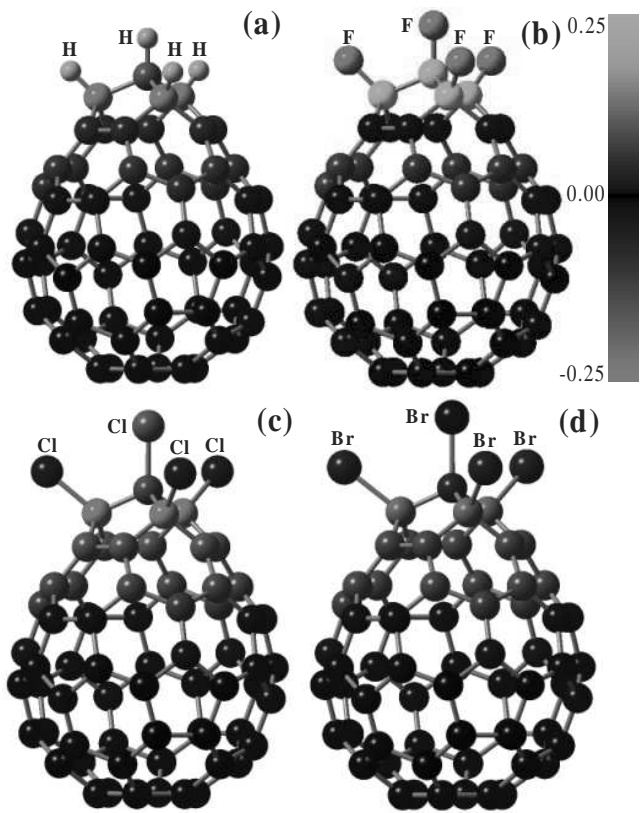}
\caption{Schematic structures and Mulliken atomic charge populations
of C$_{\texttt{3v}}$ C$_{64}$X$_{4}$ molecules: (a) C$_{64}$H$_{4}$,
(b) C$_{64}$F$_{4}$, (c) C$_{64}$Cl$_{4}$, and (d) C$_{64}$Br$_{4}$.
The capital letters indicate the exohedral atoms, and the rest atoms
are all C atoms. The color of the atoms show the Mulliken atomic
charge amount in electrons. } \label{C64X4_structure}
\end{figure}

Fig. 2 also show the Mulliken atomic charge populations of
C$_{64}$X$_{4}$ ($X=H, F, Cl, Br$). The PPFV C atoms (i.e. those
bonded with X atoms) in C$_{64}$H$_{4}$ are colored in red,
indicating that their charges are rather negative. The other atoms
at the tapering top of $C_{64}$ cage are marked in dark green,
showing that they carry little positive charges. The atoms at bottom
are electroneutral as they are in black color. The charge
populations of C$_{64}$X$_{4}$ ($X=Cl, Br$) are mainly similar to
C$_{64}$H$_{4}$. The PPFV C atoms in C$_{64}$X$_{4}$ ($X=Cl, Br$)
are also red, negative charged; while the tapering top of $C_{64}$
cage are dark green, little positive charged. Interestingly, the
colors of X atoms (namely light green, dark green, and dark red for
$X=H, Cl, Br$, respectively), are quite distinct, revealing that the
charge transfers of $H, Cl, Br$ atoms, when fused to the $C_{64}$
cage, are different. The charges of H atoms are rather positive,
while those of Cl atoms are less positive, and those of Br atoms are
even little negative. With regard to electron transfers, the PPFV C
atoms act as electron acceptors, while H and Cl atoms act as
electron donors, but Br could not. Surprisingly, the addition of F
atoms behaves rather differently. Contrary to C$_{64}$X$_{4}$ ($X=H,
Cl, Br$), the PPFV C atoms are positive charged (bright green color)
and F atoms are negative charged (light red color), showing a
totally inverse electron transfers. All the above results imply that
a proper chemical modification on PPFV could change the electronic
structures of the non-IPR carbon cage, but adding different atoms
would bring different consequences.

For C$_{64}$H$_{4}$ and C$_{64}$Cl$_{4}$, we have calculated the C-X
bond energy $\Delta{\texttt{E}}$. In the same method, the obtained
C-X bond energy $\Delta{\texttt{E}}$ for C$_{64}$X$_{4}$ ($X=F, Br$)
are 4.2810 eV, 2.4200 eV, respectively. As $\Delta{\texttt{E}}$ of
C$_{64}$H$_{4}$ is 2.6697 eV, it seems that C$_{64}$F$_{4}$ is even
more stable than C$_{64}$H$_{4}$.

We have also calculated the LUMO-HOMO gap of C$_{64}$ and
C$_{64}$X$_{4}$ ($X=H, F, Cl, Br$) by B3LYP/6-31G$^{*}$. The gap of
C$_{64}$ is 1.0879 eV, while those of C$_{64}$X$_{4}$ are 2.6697 eV,
2.7075 eV, 2.6923 eV, 2.6762 eV for $X=H, F, Cl, Br$, respectively,
which is comparable to the gap energy (2.758eV) of C$_{60}$
calculated within the same computational method. A large LUMO-HOMO
gap energy has been taken as an indication of a large stability of
fullerenes\cite{gap1,gap2}. The remarkable enlargement of the gap
energy from C$_{64}$ to C$_{64}$X$_{4}$ implies that the chemical
deriving could indeed enhance the stability of non-IPR fullerenes.

\begin{figure}[tb]
\includegraphics[width=5.4in, keepaspectratio]{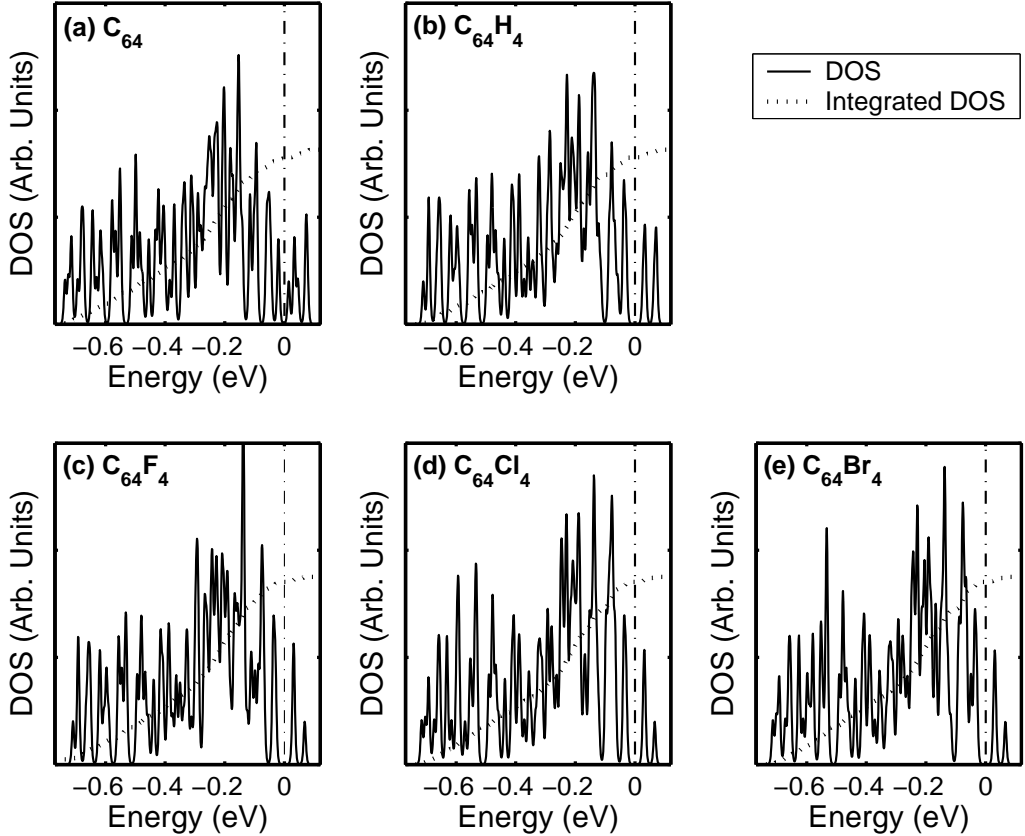}
\caption{Density of states (DOS) and the integrated DOS(doted line)
of electrons as a function of energy for C$_{64}$ and
C$_{64}$X$_{4}$ ($X=H, F, Cl, Br$). The energy zero-point is taken
at the Fermi level, that is indicated by the vertical dash-dot
line.} \label{C64X4_DOS}
\end{figure}

The DOS provide a convenient overall view of the electronic
structure of clusters and solids. Fig. 3 exhibits the DOS of
C$_{64}$ and C$_{64}$X$_{4}$ ($X=H, F, Cl, Br$), which are obtained
by Fermi-Dirac smearing of molecular energy levels. We can see that
from C$_{64}$ to C$_{64}$X$_{4}$ ($X=H, F, Cl, Br$), LUMO-HOMO gaps
are enlarged.


To provide a verifying basis for the experimental identification of
C$_{64}$X$_{4}$ ($X=H, F, Cl, Br$), we have calculated the IR, Raman
spectra by B3LYP/6-31G$^{*}$, and NMR spectra by
GIAO-B3LYP/6-31G$^{*}$, with the optimized geometry in the
B3LYP/6-31G$^{*}$ level, as shown in Figs. 4 and 5.

\begin{figure}[tb]
\includegraphics[width=3.6in, keepaspectratio]{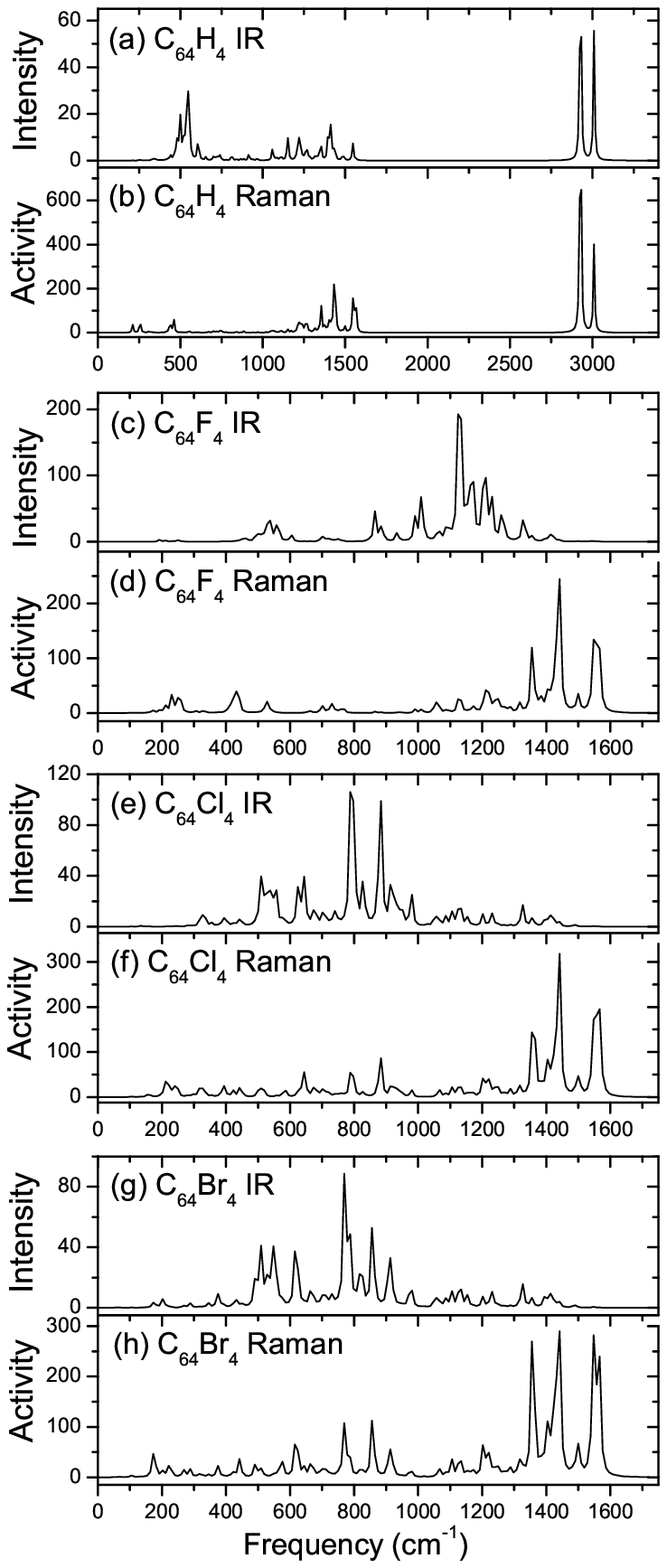}
\caption{IR and Raman spectra of C$_{\texttt{3v}}$ C$_{64}$X$_{4}$
(X=H, F, Cl, Br), which are scaled by a factor of 0.9613.}
\label{C64X4_IR_Raman}
\end{figure}

As can be seen in Figs. 4(a) and 4(b), the computed IR spectrum of
C$_{64}$H$_{4}$ agrees well with the experimental
results\cite{Wang}; both IR and Raman spectra of C$_{64}$H$_{4}$
spread up to more than 3000 cm$^{-1}$ and have two major regions.
The region of two sharp peaks near 3000 cm$^{-1}$ corresponds to the
C-H stretching modes which are rather active in both IR and Raman
spectra, and the wide region in the range less than 1700 cm$^{-1}$
corresponds to the C-C stretching, C-C-C bending and C-C-H bending
modes in which active modes are different for IR and Raman spectra.

The IR and Raman spectra of C$_{64}$X$_{4}$ ($X=F, Cl, Br$) are
rather different, as depicted in Figs. 4(c) to 4(h). Unlike
C$_{64}$H$_{4}$, they are only extended to circa 1700 cm$^{-1}$ and
have no distinct separated regions. From a view of vibrational
modes, the spectra of C$_{64}$X$_{4}$ ($X=F, Cl, Br$) lack a clear
region of C-X stretching modes like C$_{64}$H$_{4}$. It may be
because H atom is much lighter than C atom, the C-H stretching modes
could be separated from other modes caused by C atoms, and the
corresponding peaks are located at a very high frequency region. By
contrast, F, Cl and Br atoms are heavier than C atom, the motion of
F, Cl and Br atoms must be coupled with C atoms strongly, resulting
in that the IR and Raman spectra of C$_{64}$Cl$_{4}$ could not be
parted into two regions. This may also cause other spectral
differences between C$_{64}$H$_{4}$ and C$_{64}$Cl$_{4}$.

For C$_{64}$X$_{4}$ ($X=F, Cl, Br$), the IR spectra profiles exhibit
some similarity, but the main peaks are located at different
frequencies. It seems that, from C$_{64}$F$_{4}$ to
C$_{64}$Br$_{4}$, the main peaks of IR spectra are red shifted, and
intensities are reduced. Raman spectra show more similarity: not
only are the profiles very similar, but also the main peaks are all
located between 1400 cm$^{-1}$ and 1600 cm$^{-1}$. By a careful
check, the Raman spectrum of C$_{64}$H$_{4}$ below 1600 cm$^{-1}$ in
Fig. 4(b) also presents the above similarity to that of
C$_{64}$X$_{4}$ ($X=F, Cl, Br$). In addition, the Raman spectral
activity grows from C$_{64}$H$_{4}$ to C$_{64}$Br$_{4}$. We can
conclude that the IR spectra are more sensitive to different added X
atoms than Raman spectra: the heavier added X atoms will red-shift
the main active IR modes and reduce their intensities, but will
enhance the activity of Raman modes and not affect the active Raman
modes frequencies.

\begin{figure}[tb]
\includegraphics[width=3.6in, keepaspectratio]{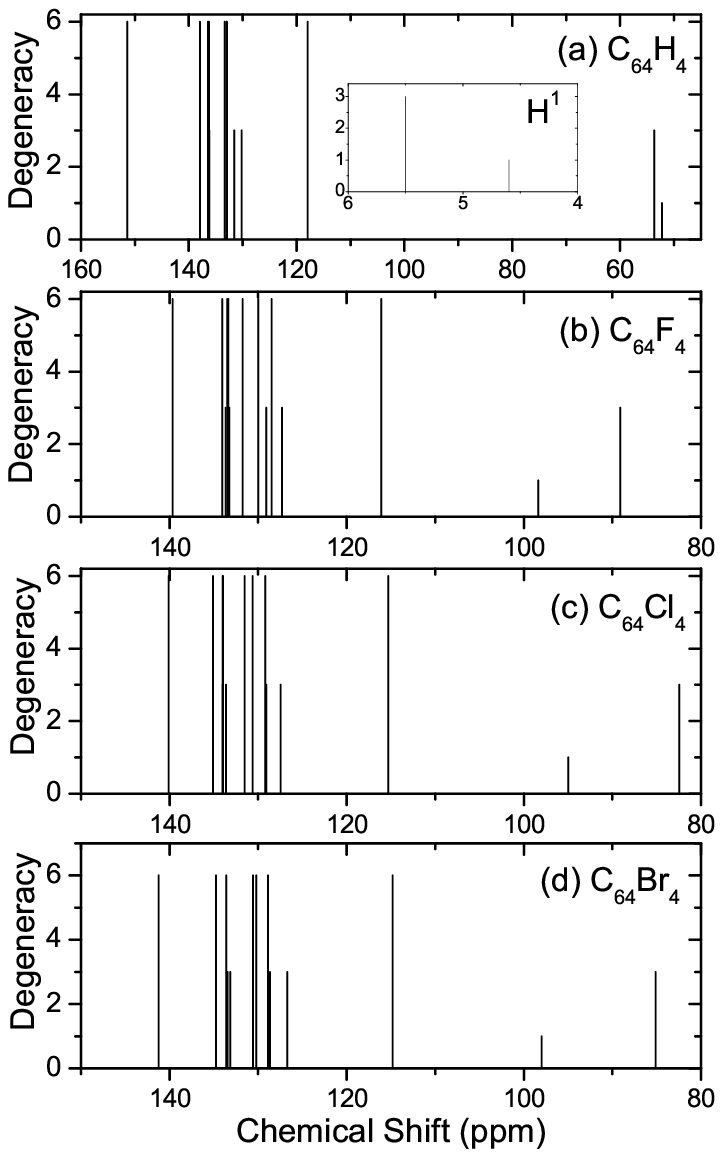}
\caption{NMR spectra of C$_{\texttt{3v}}$ C$_{64}$X$_{4}$ ($X=H, F,
Cl, Br$), scaled by referenced shifts, where the inset of (a) shows
the chemical shifts of H atoms, others show that of C atoms.}
\label{C64X4_NMR}
\end{figure}

Fig. 5 gives the calculated NMR spectra of C$_{64}$X$_{4}$ ($X=H, F,
Cl, Br$). The computed NMR spectrum of C$_{64}$H$_{4}$ agrees well
with the experimental results as depicted in Fig. 4(a). The
C$_{\mathtt{3v}}$ C$_{64}$X$_{4}$ have 14 unique types of C atoms
[60 $sp^{2}$ C atoms $(8\times6;4\times3)$; 4 $sp^{3}$ C atoms
$(1\times3;1\times1)$] and 2 unique types of X atoms [total 4 X
atoms $(1\times3;1\times1)$], where $a\times b$ represents $a$
unique types, and in one type, there are $b$ symmetrical equivalent
atoms. All of the calculated C$^{13}$ NMR spectra of C atoms in
C$_{64}$X$_{4}$ ($X=H, F, Cl, Br$) are easily divided into two
regions. One may see that there are 12 peaks in the left region, 8
of which have the same intensity, and the other 4 have the half of
the intensity, which could be ascribed to the
$sp^{2}(8\times6;4\times3)$ C atoms. The right regions of
C$_{64}$X$_{4}$ ($X=H, F, Cl, Br$) only have two peaks, whose
relative intensities are 3:1. It is interesting to note that the two
peaks in C$_{64}$H$_{4}$ are very close, where the higher peak is in
the left, whereas the two peaks in C$_{64}$F$_{4}$ are separated
from each other, where the higher peak is in the right. We can
ascribe this feature to the $sp^{3}(1\times3;1\times1)$ C atoms
which are bonded to the X atoms (i.e. PPFV). The lower peak could be
ascribed to the single C atom on the $C_{3}$ axis, and another peak
could be ascribed to the equivalent triple C atoms out of the
$C_{3}$ axis. The NMR spectra of C$_{64}$Cl$_{4}$ and
C$_{64}$Br$_{4}$ are rather similar to those of C$_{64}$F$_{4}$,
with small quantitative differences. The calculated H$^{1}$ NMR
spectrum of H atoms in C$_{64}$H$_{4}$ is also included, where the
two peaks are observed in an intensity ratio of about 3:1.
Obviously, the lower peak can be ascribed to the single H atom on
the $C_{3}$ axis, whereas another peak can be ascribed to the
equivalent triple H atoms out of the $C_{3}$ axis.


\begin{figure}[tb]
\includegraphics[width=6.0in,clip]{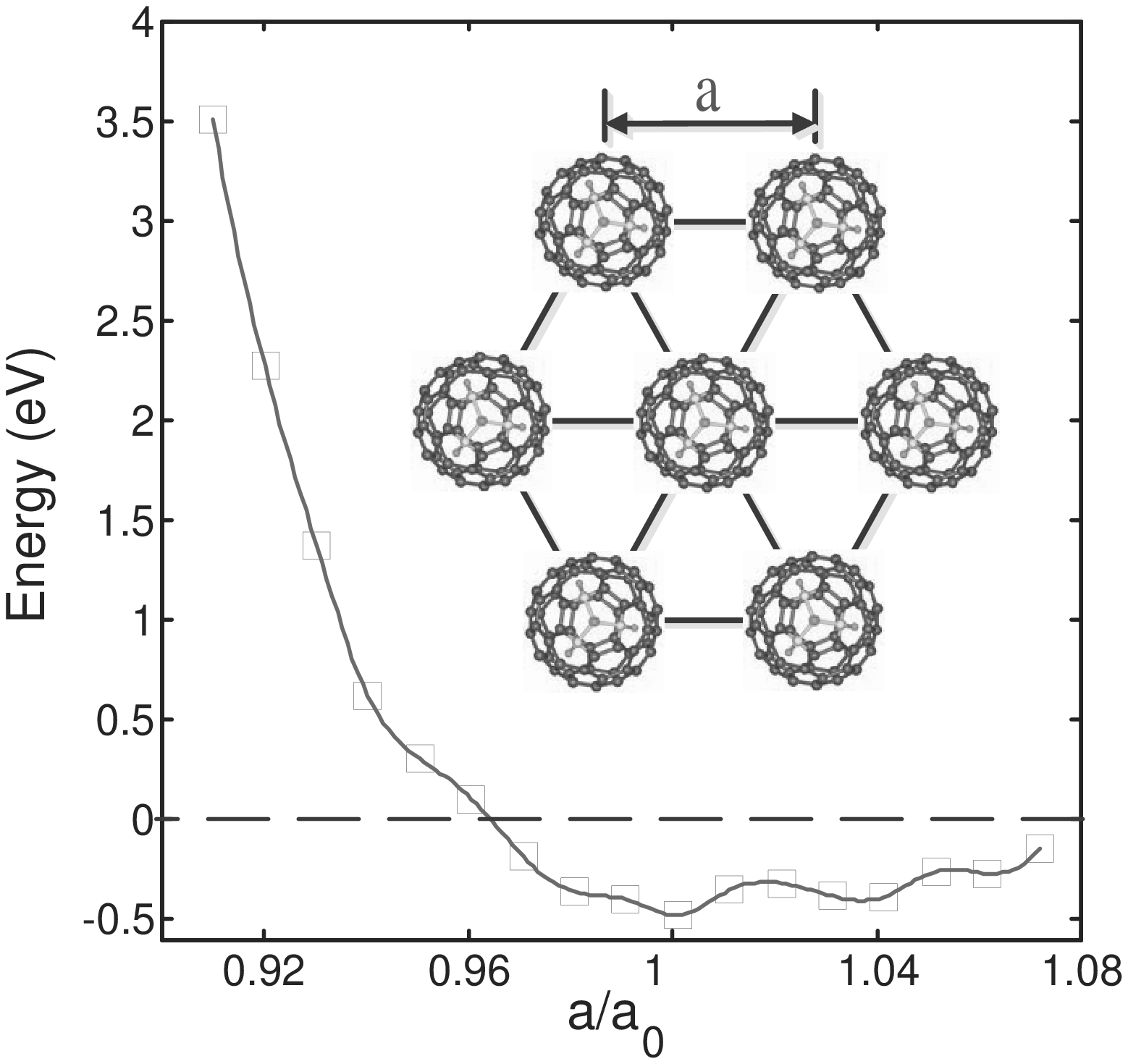}
\caption{The total energy per C$_{64}$H$_{4}$ molecule in hexagonal
monolayer, measured from the total energy of the isolated
C$_{64}$H$_{4}$ molecule, as a function of the lattice constant,
where the optimized lattice constant, $a_{0}=$9.89
${\mathnormal{\mathring{A}}}$.} \label{monolayer}
\end{figure}

Let us now address another interesting issue if C$_{64}$X$_{4}$
($X=H, F, Cl, Br$) molecules can condense to form a monolayer
structure. We have considered the hexagonal lattice, which
has been simulated by a supercell with layer-layer distance of 30
${\mathnormal{\mathring{A}}}$. The C$_{64}$X$_{4}$ molecules are put
on the lattice site with $C_{3}$ axis perpendicular to the monolayer
plane. Then, the total energy is calculated accordingly with respect
to different lattice parameters (every 0.1
${\mathnormal{\mathring{A}}}$) for the presumed structure. For
C$_{64}$H$_{4}$ hexagonal monolayer, as indicated in Fig. 6, a
clearly single minimal total energy with the optimized lattice
constant $ a_{0}=9.89$ ${\mathnormal{\mathring{A}}}$ is obtained,
revealing that this structure is stable and easily to form. The data
with 0.01 ${\mathnormal{\mathring{A}}}$ precision are obtained by a
spline interpolation from the calculated data with 0.1
${\mathnormal{\mathring{A}}}$ precision. The cohesive energy per
C$_{64}$H$_{4}$ molecule for this monolayer is found to be 0.48 $
eV$. For C$_{64}$X$_{4}$ ($X=F, Cl, Br$), the hexagonal monolayer
structure is also found to be stable, and the optimized lattice
constants are still $ a_{0}=9.89$ ${\mathnormal{\mathring{A}}}$. The
cohesive energy per molecule are 0.67 eV, 0.55 eV, 0.32 eV,
respectively. It appears that C$_{64}$F$_{4}$ hexagonal monolayer phase is
the most stable among them.


To summarize, in terms of first-principles calculations based on the
density functional theory, we have systematically reported the
structures, electronic properties and spectroscopies of the unconventional
fullerene C$_{64}$ and its derivatives C$_{64}$X$_{4}$
($X=H,F,Cl,Br$). We have found that the parent carbon cage of the recently
synthesized C$_{64}$H$_{4}$ with C$_{\texttt{3v}}$ symmetry is not the
ground state of C$_{64}$, while the latter should be a spherical shaped, D$_{\texttt{2}}$
symmetrical isomer with spiral number 3451. Our computation results
show that the combination of the four X ($X=H, Cl, F, Br$) atoms to the
pentagon-pentagon fusion vertex of C$_{64}$ cage could change the
electronic structures and enhance the stability of the fullerenes, and C$_{64}$X$_{4}$
($X=F,Cl,Br$) may be also stable unconventional fullerides like C$_{64}$H$_{4}$. Similar to
C$_{64}$H$_{4}$, C$_{64}$X$_{4}$ ($X=F,Cl,Br$) are found to be
pear-shaped, with large LUMO-HOMO gap energies and large C-X bond
energy. It appears that C$_{64}$F$_{4}$ is even more stable than
C$_{64}$H$_{4}$ as the C-X bond energy of the former is higher than
that of the latter. The computed spectroscopies of C$_{64}$H$_{4}$ molecule agree
well with the experimental data; the IR, Raman, NMR spectra of
C$_{64}$X$_{4}$ ($X=F,Cl,Br$) are also presented to stimulate further
experimental investigations. In addition, it is revealed that, by
total energy calculations, C$_{64}$X$_{4}$ could form stable
hexagonal monolayers and among them, C$_{64}$F$_{4}$ hexagonal monolayer
is the most stable.

\noindent
 The authors are grateful to B. Gu, H. F. Mu, and Z. C. Wang
for helpful discussions. All of the calculations are completed on
the supercomputer NOVASCALE 6800 in Virtual Laboratory of
Computational Chemistry, Computer NetWork Information Center
(Supercomputing Center) of Chinese Academy of Sciences. This work is
supported in part by the National Science Foundation of China (Grant
Nos. 90403036, 20490210), and by the MOST of China (Grant No.
2006CB601102).



\begin{thebibliography}{99}
\bibitem{c60discover} Kroto, H. W.; Heath, J. R.; O'Brien, S. C.; Curl, R. F.; Smalley, R. E. \emph{Nature} \textbf{1985}, \emph{318}, 162.

\bibitem{IPR1} Kroto, H. W. \emph{Nature} \textbf{1987}, \emph{329}, 529.

\bibitem{IPR2} Kadish, K. M.; Ruoff, R. S.; Editors, \textit{Fullerene:
Chemistry, Physics and Technology}; Wiley: New York, 2002.

\bibitem{minimum_curvaure} Albertazzi, E.; Domene, C.; Fowler, P.
W.; Heine, T.; Seifert, G.; Alsenow, C. V.; Zerbetto, F. \emph{Phys.
Chem. Chem. Phys} \textbf{1999}, \emph{1}, 2913.

\bibitem{stablefuller} Taylor, R.; Avent, A. G.; Birkett, P. R., Dennis, T. J. S.; Hare, J. P.; Hitchcock, P. B.; Holloway, J. H.; Hope, E. G.; Kroto, H. W.; Langley, G. J.; Meidine, M. F.; Parsons, J. P.; Walton, D. R. M. \emph{Pure \& Appl. Chem.}, \textbf{1993}, \emph{65}, 135.

\bibitem{endohedral_CRWang} Wang, C.-R.; Kai, T.; Tomiyama, T.; Yoshida, T.; Kobayashi, Y.; Nishibori, E.; Takata, M.; Sakata, M.; Shinohara, H. \emph{Nature} \textbf{2000} , \emph{408}, 426.

\bibitem{endohedral_Stevenson} Stevenson, S.; Fowler, P. W.; Heine, T.; Duchamp, J. C.; Rice, G.; Glass, T.; Harich, K., Hajdu, E.; Bible, R.; Dorn, H. C. \emph{Nature} \textbf{2000}, \emph{408}, 427.

\bibitem{endohedral_Kato}Kato, H.; Taninaka, A.; Sugai, T.; Shinohara, H. \emph{J.
Am. Chem. Soc.} \textbf{2003}, \emph{125}, 7782.

\bibitem{Xie} Xie, S.-Y.; Gao, F.; Lu, X.; Huang, R.-B.; Wang, C.-R.; Zhang, X.; Liu, M.-L.; Deng, S.-L.; Zheng, L.-S. \emph{Science} \textbf{2004}, \emph{304}, 699.

\bibitem{Wang}Wang, C.-R.; Shi, Z.-Q.; Wan, L.-J.; Lu, X.; Dunsch, L.; Shu, C.-Y.; Tang, Y.-L.; Shinohara, H. \emph{J. Am. Chem. Soc.} \textbf{2006}, \emph{128},
6605.

\bibitem{C50Cl10monolayer} Yan, Q.-B.; Zheng, Q.-R.; Su, G. \emph{Phys. Rev. B}
\textbf{2006}, \emph{73}, 165417.

\bibitem{spiral} Fowler, P. W.;  Manolopoulos, D. E. \emph{An Atlas of
Fullerenes}; Clarendon: Oxford, 1995.

\bibitem{MNDO} Dewar, M. J. S.; Thel, W. \emph{J. Am. Chem. Soc.} \textbf{1977},
\emph{99}, 4899.

\bibitem{semiempirical} Chen, Z.; Thiel, W.  \emph{Chem. Phys.
Lett.} \textbf{2003}, \emph{367}, 15.

\bibitem{C40PAPR} Albertazzi, E.; Domene, C.; Fowler, P. W.; Heine, T.; Seifert,
G.; Alsenow, C. V.; Zerbetto, F. \emph{Phys. Chem. Chem. Phys.}
\textbf{1999}, \emph{1}, 2913.

\bibitem{Hohenberg-Kohn}Hohenberg, P.; Kohn, W. \emph{Phys. Rev.}
\textbf{1964}, \emph{136}, B864.

\bibitem{B3} Becke, A. D. \emph{J. Chem. Phys.} \textbf{1993}, \emph{98}, 5648.

\bibitem{LYP} Lee, C.; Yang, W.; Parr, R. G. \emph{Phys. Rev. B}
\textbf{1988}, \emph{37}, 785.

\bibitem{Gaussian03} Gaussian 03, Revision C.02,
Frisch, M. J.; Trucks, G. W.; Schlegel, H. B.; Scuseria, G. E.;
Robb, M. A.; Cheeseman, J. R.; Montgomery, J. A.; T. Vreven, Jr.;
Kudin, K. N.; Burant, J. C.; Millam, J. M.; Iyengar, S. S.;
Tomasi,J.; Barone, V.; Mennucci, B.; Cossi, M.; Scalmani, G.; Rega,
N.; Petersson, G. A.; Nakatsuji, H.; Hada, M.; Ehara, M.; Toyota,
K.; Fukuda, R.; Hasegawa, J.; Ishida, M.; Nakajima, T.; Honda, Y.;
Kitao, O.; Nakai, H.; Klene, M.; Li, X.; Knox, J. E.; Hratchian, H.
P.; Cross, J. B.; Adamo, C.; Jaramillo, J.; Gomperts, R.; Stratmann,
R. E.; Yazyev, O.; Austin, A. J.; Cammi, R.; Pomelli, C.; Ochterski,
J. W.; Ayala, P. Y.; Morokuma, K.; Voth, G. A.; Salvador, P.;
Dannenberg, J. J.; Zakrzewski, V. G.; Dapprich, S.; Daniels, A. D.;
Strain, M. C.; Farkas, O.; Malick, D. K.; Rabuck, A. D.;
Raghavachari, K.; Foresman, J. B.; Ortiz, J. V.; Cui, Q.; Baboul, A.
G.; Clifford, S.; Cioslowski, J.; Stefanov, B. B.; Liu, G.;
Liashenko, A.; Piskorz, P.; Komaromi, I.; Martin, R. L.; Fox, D. J.;
Keith, T.; Al-Laham, M. A.; Peng, C. Y.; Nanayakkara, A.;
Challacombe, M.; Gill, P. M. W.; Johnson, B.; Chen, W.; Wong, M. W.;
Gonzalez, C.; Pople, J. A.
 Gaussian, Inc., Wallingford CT, 2004.

\bibitem{ABINIT} The ABINIT code is a common project of the Universite
Catholique de Lovain, Corning Incorporated, and other contributors
(http://www.abinit.org).

\bibitem{NCPP}Troullier, N.;  Martins, J. L. \emph{Phys. Rev. B}
\textbf{1991}, \emph{43}, 1993.

\bibitem{Teter}Goedecker, S.; Teter, M.; Hutter, J. \emph{Phys. Rev. B} \textbf{1996
}, \emph{54}, 1703.

\bibitem{C50_XinLu} Lu, X.; Chen, Z.; Thiel, Wa.; Schleyer, P. von
R.; Huang, R.; Zheng,  L. \emph{J. Am. Chem. Soc.}
\textbf{2004},\emph{126}, 14871.

\bibitem{Delta_E}Chiu, Y. N.; Ganelin, P.; Jiang, X.; Wang, B. C.;  \emph{J. Mol. Struc.
(Theochem)} \textbf{1994}, \emph{312}, 215.

\bibitem{exhydrogenate_order} Liu, M.; Chiu, Y.-N.; Xiao, J. \emph{J. Mol. Struc. (Theochem)} \textbf{1999}, \emph{489},
109.

\bibitem{gap1}Manopoulos, D. E.; May, J. C.; Down, S. E.;  \emph{Chem. Phys. Lett.}
\textbf{1991}, \emph{181}, 105.

\bibitem{gap2}Diener, M. D.; Alford, J. M.  \emph{Nature} \textbf{1998}, \emph{393}, 668.
\end{thebibliography}
\end{document}